\documentclass[aps,prb,amsmath,amssymb,twocolumn,superscriptaddress,floatfix]{revtex4-1}
\usepackage{graphicx}
\usepackage{helvet}
\usepackage{hyperref}
\hypersetup{pdftitle={Armchair-Kagome Mn2Sb2O7},pdfauthor={Darren C. Peets},pdfsubject={Frustrated magnetism},pdfdisplaydoctitle}

\def\Mn{Mn\ensuremath{_2}Sb\ensuremath{_2}O\ensuremath{_7}}
\def\py{pyr-Mn\ensuremath{_2}Sb\ensuremath{_2}O\ensuremath{_7}}
\def\TN1{\ensuremath{T_\text{N1}}}
\def\TN2{\ensuremath{T_\text{N2}}}
\def\TCW{\ensuremath{T_\text{CW}}}
\def\P3{\ensuremath{P3_121}}
\begin{document}

\title{Magnetic Transitions in the Chiral Armchair-Kagome System Mn$_2$Sb$_2$O$_7$}

\author{Darren C. Peets}
\email{dpeets@fudan.edu.cn}
\altaffiliation[Current address: ]{Advanced Materials Laboratory, Fudan University, Shanghai 200438, China}
\author{Hasung Sim}
\author{Seongil Choi}
\affiliation{Center for Correlated Electron Systems, Institute for Basic Science (IBS), Seoul 08826, Korea}
\affiliation{Department of Physics and Astronomy, Seoul National University, Seoul 08826, Korea}

\author{Maxim Avdeev}
\affiliation{Australian Nuclear Science and Technology Organisation, Lucas Heights, NSW 2234, Australia}

\author{Seongsu Lee}
\author{Su Jae Kim}
\affiliation{Neutron Science Division, Korea Atomic Energy Research Institute, Daejeon 34057, Korea}

\author{Hoju Kang}
\author{Docheon Ahn}
\affiliation{Beamline Department, Pohang Accelerator Laboratory, 80 Jigokro-127-beongil, Nam-gu, Pohang 37673, Gyeongbuk, Korea}

\author{Je-Geun Park}
\email{jgpark10@snu.ac.kr}
\affiliation{Center for Correlated Electron Systems, Institute for Basic Science (IBS), Seoul 08826, Korea}
\affiliation{Department of Physics and Astronomy, Seoul National University, Seoul 08826, Korea}

\date{\today}

\begin{abstract}
  The competition between interactions in frustrated magnets allows a
  wide variety of new ground states, often exhibiting emergent physics
  and unique excitations.  Expanding the suite of lattices available
  for study enhances our chances of finding exotic physics.
  \Mn\ forms in a chiral, kagome-based structure in which a fourth
  member is added to the kagome-plane triangles to form an armchair
  unit and link adjacent kagome planes.  This structural motif may be
  viewed as intermediate between the triangles of the kagome network
  and the tetrahedra in the pyrochlore lattice.  \Mn\ exhibits two
  distinct magnetic phase transitions, at 11.1 and 14.2\,K, at least
  one of which has a weak ferromagnetic component.  The magnetic
  propagation vector does not change through the lower transition,
  suggesting a metamagnetic transition or a transition involving a
  multi-component order parameter.  Although previously reported in
  the $P3_121$ space group, \Mn\ actually crystallizes in $P2$, which
  allows ferroelectricity, and we show clear evidence of
  magnetoelectric coupling indicative of multiferroic order.  The
  quasi-two-dimensional `armchair-kagome' lattice presents a promising
  platform for probing chiral magnetism and the effect of
  dimensionality in highly frustrated systems.
\end{abstract}

\maketitle

\section{Introduction}

In frustrated magnetic systems, the competition among interactions
leads to a rich and sometimes tuneable array of physics, which can be
quite subtle and sensitive to perturbations\cite{Lacroix2011}.  The
large and degenerate Hilbert space created in such systems lends
itself to a multitude of magnetic ground states and low-lying excited
states.  Generally, frustration can be introduced {\it via} competing
interactions, such as a balance of direct exchange, superexchange, and
Dzyaloshinskii-Moriya interactions, or geometrically, by arranging the
magnetic ions in a lattice that prevents a pairwise
(anti)ferromagnetic ground state from being simultaneously satisfied
within all nearest-neighbour pairs of magnetic atoms.  Several such
lattices such as pyrochlore, kagome, and triangular are well-known.
When new variants of established magnetic lattices are reported, {\it
  e.g.} the hyperhoneycomb or tripod-kagome
lattices\cite{Takayama2015,Dun2016}, they offer exciting new
playgrounds for the investigation of frustration.

Chiral magnetism in particular offers the possibilities of highly
nontrivial long-range magnetic order and topological excitations such
as skyrmions\cite{Skyrmions}.  While competing interactions or weak
effects such as the Dzyaloshinskii-Moriya
interaction\cite{Dz1958,Moriya1960} can lead to chiral magnetic
structures even in the case of an achiral crystal lattice, it may be
more natural to expect chiral magnetism to arise if the underlying
lattice is itself chiral.  The magnetic order may be expected to
follow the symmetry of the lattice, even in the absence of the
delicate balance of interactions otherwise required, and in many cases
all irreducible representations for the magnetic order will be chiral,
{\slshape constraining} the magnetic order to be chiral by symmetry.
Here we highlight one magnetic material with a novel frustrated
lattice in just such a chiral crystal structure.

\begin{figure}[thb]
\includegraphics[width=\columnwidth]{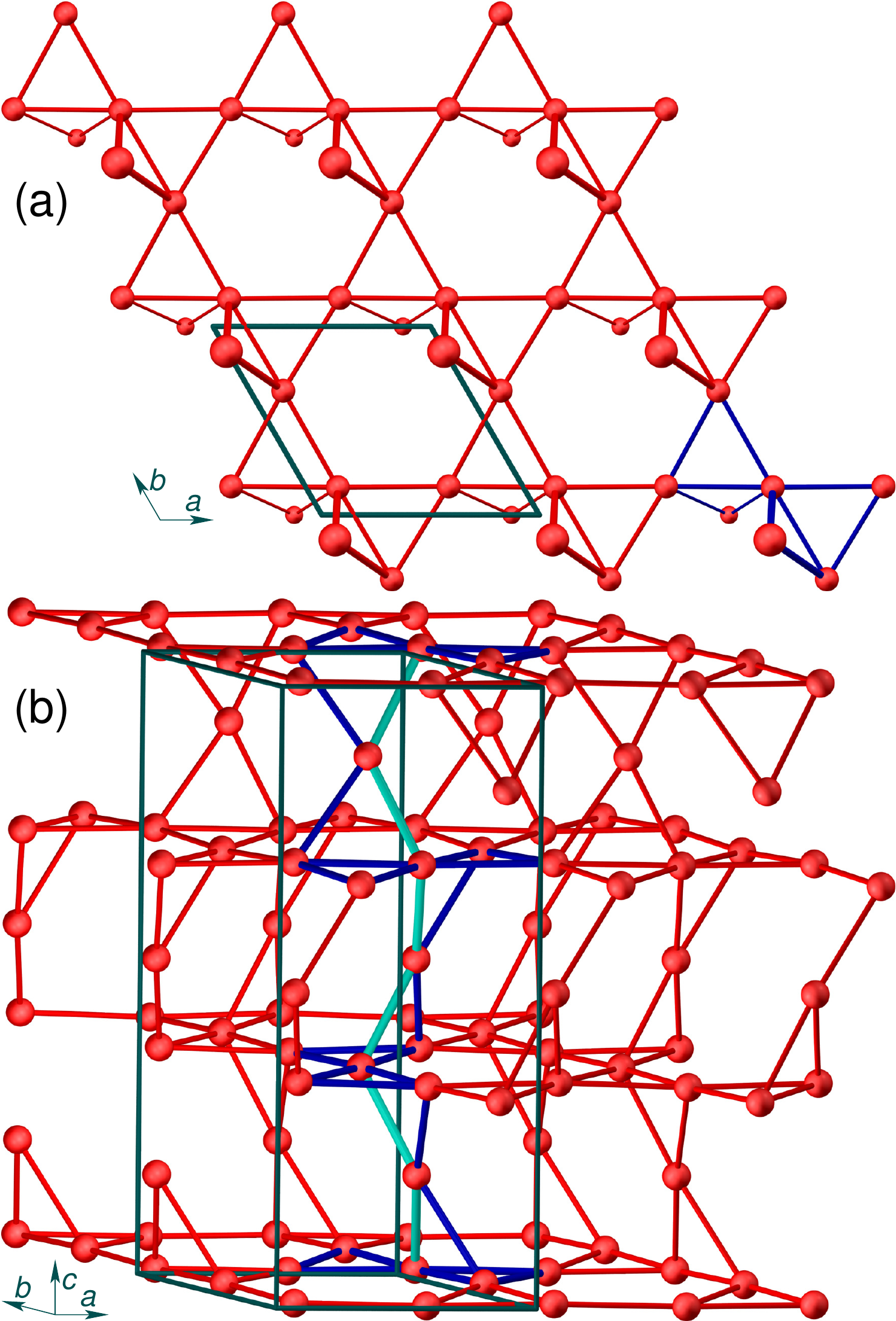}
\caption{\label{fig:P3121}Manganese sublattice in the \P3\ structure
  of \Mn\ at 600$^\circ$C, highlighting the `armchair-kagome' network,
  based on the refinement in Fig.~\ref{fig:600} and
  Tab.~\ref{tab:600C}.  (a) Network of armchair units at $z=0$; one
  pair of adjacent armchairs is highlighted in blue.  (b) The layer
  stacking leads to a 3-fold screw axis.  The stacking sequence of
  armchair units around the rear corner of the \P3\ unit cell is shown
  in blue, with the shortest bonds in lighter blue.  The unit cell is
  shown using solid lines.}
\end{figure}

When prepared by standard high-temperature solid-state synthesis,
\Mn\ forms in a distorted, chiral variant of the trigonal Weberite
structure, as shown in
Fig.~\ref{fig:P3121}\cite{Scott1987,Scott1990,Chelazzi2013}.  The
material's Mn network is composed of slightly distorted kagome layers
in the $ab$ plane, with the kagome triangles linked along the $c$ axis
through additional Mn triangle units, such that the simplest
structural motif to consider is an Mn$_4$ armchair unit.  Here, we
neglect the Sb sublattice, which is known to be nonmagnetic Sb$^{5+}$
from $^{121}$Sb M\"ossbauer
spectroscopy\cite{Subramanian1984}\footnote{Ref.~\onlinecite{Subramanian1984}
  describes the crystal structure as rhombohedrally-distorted
  pyrochlore --- the correct space group had not yet been identified.
  The samples were prepared by a high-temperature route that produces
  the trigonal Weberite structure investigated
  here\cite{Reimers1991}.}.  The shortest Mn--Mn bonds in our
refinement trace out a helix along the $c$ axis, highlighted in light
blue.  In the case of a pyrochlore lattice, the atom forming the
vertical link would lie directly over the centre of a kagome-layer
triangle instead of off to one side, forming a tetrahedron and leading
to a highly-symmetric structure in which the faces of the tetrahedra
comprise interpenetrating kagome networks.  Armchair-kagome \Mn\ is
lower-dimensional, but the interactions both along the $c$ axis and
in-plane should exhibit geometric frustration due to the triangular
arrangements of Mn atoms.  It would be possible, in principle, to
deform the Mn sublattice in \Mn\ into a pyrochlore network through a
series of slips perpendicular to the $c$-axis, but each slip would
occur along an entire plane the size of the crystal, and six slips
would be required per unit cell, making such a reconstruction
enormously energetically unfavorable.  This would be a topological
transition, since there is one more Mn--Mn link in the pyrochlore's
{\slshape T}$_4$ tetrahedron than in an Mn$_4$
armchair\cite{Kane2013}.  The pyrochlore polymorph of \Mn, \py, also
exists but can only be prepared by a more-difficult low-temperature
route\cite{Brisse1972,Zhou2008,Zhou2010,Peets-pyr}.  In this paper,
`\Mn' refers to the armchair-kagome polymorph unless otherwise stated.

In terms of physical properties, little has been reported on
\Mn\ to date.  It has a paramagnetic Curie-Weiss temperature
\TCW\ around $-45$ to $-50$\,K with a paramagnetic moment
corresponding to high-spin $3d^5$ Mn$^{2+}$,\cite{Reimers1991} and it
undergoes a bulk magnetic transition around 13\,K.  The clear history
dependence reported below about 55\,K was attributed to an apparently
abrupt onset of short-range correlations.  The magnetic structure has
not been solved, but most magnetic Bragg peaks could be explained by a
[$\frac{1}{2}$00] propagation vector.  $^{121}$Sb M\"ossbauer spectra
have been reported\cite{Subramanian1984}, but no splitting of the
sites was observed, nor any Sb$^{3+}$ component, nor hyperfine
splitting in the magnetically-ordered phase.  The material's
isothermal bulk modulus is also known\cite{Chelazzi2013}, but the
authors are aware of no other measurements.

We set out to clarify the crystal structure and the nature of the
magnetic order in \Mn, and to check for hints of exotic behavior
arising from its unique frustrated chiral lattice.  The material forms
in an even lower-symmetry structure than that previously reported, and
the magnetic order is frustrated, chiral, and multiferroic.
\Mn\ first enters a magnetically-ordered state around 14\,K in which
the magnetization increases rapidly on cooling, followed by a second
magnetic transition below which the magnetization saturates.

\section{Experimental}

Powder samples of \Mn\ were prepared in Al$_2$O$_3$ crucibles in air,
from intimately mixed MnO$_2$ (Alfa Aesar, 99.997\%) and Sb$_2$O$_3$
(Alfa Aesar, 99.999\%).  Mixed powders were calcined with intermediate
grindings at temperatures between 1050 and 1150$^\circ$C, typically
for 24\,h per temperature; the mass was monitored for loss of volatile
component oxides, and powder diffraction patterns were used to verify
phase purity.  X-ray powder patterns reported here were collected at
temperatures from 30 to 1000$^\circ$C using a Bruker D8 Discover
diffractometer with a Cu$K\alpha$ source.  When higher-density samples
were desirable, most notably for specific heat and dielectric constant
measurements, this powder was pressed isostatically into rods and
sintered at 1175$^\circ$C for a further 24\,h to produce ceramic.

Magnetization measurements in fields up to 5\,T were performed in a
Quantum Design MPMS-XL SQUID magnetometer in its RSO measurement mode.
A powder sample of approximately 15\,mg was packed inside a gelatin
capsule, which was closed with Kapton tape and loaded into a plastic
straw.  The contribution from the empty sample holder was below the
level of the noise on the \Mn\ data.  AC susceptibility measurements
were performed in the same magnetometer using the DC sample transport,
at zero applied field.  Magnetization $M(H)$ data in fields up to
14\,T were measured using a Quantum Design PPMS with the vibrating
sample magnetometry option.  Specific heat measurements were performed
by the relaxation time method in fields up to 9\,T in a Quantum Design
PPMS, with 2$\tau$ fitting and measurement times of 2$\tau$.  Sintered
ceramic samples were attached to the sample stage using Apiezon N
Grease for low-temperature measurements, while Apiezon H Grease was
used above $\sim210$\,K to avoid artifacts from the N Grease glass
transition.  The addenda contribution was subtracted within the
software.

Dielectric measurements were performed at 1\,kHz using an
Andeen-Hagerling AH2550A capacitance bridge.  A $3\times3$\,mm$^2$,
2\,mm-thick slab of \Mn\ ceramic was affixed to a gold backing plate
with silver epoxy; a copper wire connected to silver epoxy spread over
the opposite surface completed a capacitor with the \Mn\ ceramic as
its dielectric.  This assembly was mounted onto a closed-cycle
refrigerator, cooled to $\sim$5\,K, then measured on warming back to
room temperature while the CCR was turned off to reduce noise.

X-ray diffraction found the powder samples to be phase pure for up to
$\sim3$\,\% initial cation nonstoichiometry, after two or more
calcines above 1050$^\circ$C.  Magnetization measurements, however,
proved quite sensitive to the trace presence of Mn$_3$O$_4$, which has
a broad ferrimagnetic transition below
50\,K\cite{Dwight1960,Jensen1974}.  This high sensitivity enabled
better optimization of the synthesis conditions than was possible with
x-ray diffraction alone.  Ultimately, a $\sim$2\,\% antimony excess
was used.  Samples on which magnetization is presented were
additionally washed in citric acid to ensure the complete absence of
Mn$_3$O$_4$, although this was later found to be unnecessary.
\Mn\ was also found to be stable in dilute nitric and hydrochloric
acids.

A high-resolution synchrotron x-ray powder diffraction pattern was
collected at room temperature on beamline 9B (HRPD) at the Pohang
Accelerator Laboratory, in Pohang, Korea.  A specimen of approximately
0.2\,g was prepared by a flat plate side loading method to avoid
preferred orientations, and the sample was rotated about the surface
normal during the measurement to increase sampling statistics.  Data
were collected from 10 to 130.5$^\circ$ in steps of 0.005$^\circ$,
using a wavelength of 1.4970(1)\,\AA, and they were normalized to the
incoming beam intensity and corrected for asymmetric diffraction.

For higher sensitivity to oxygen atoms and to access the magnetic
structure, powder neutron diffraction was performed at a variety of
temperatures at the ECHIDNA diffractometer at the OPAL research
reactor at ANSTO, Australia, from 6.5 to 163.95$^\circ$ in steps of
0.05$^\circ$, with a neutron wavelength of 2.4395\,\AA.
Low-temperature measurements were performed on loose powder sealed in
a vanadium can, while for temperatures above room temperature, the
sample was a sintered rod suspended in vacuum.  Additional powder
neutron diffraction patterns were collected at several temperatures at
the high-resolution powder diffractometer (HRPD) at the HANARO
research reactor in Daejeon, Korea, from 0 to 159.95$^\circ$ in steps
of 0.05$^\circ$, with a neutron wavelength of 1.8343\,\AA.  To examine
the crystal structure with the highest possible resolution,
time-of-flight powder neutron diffraction data were also collected at
room temperature on the High Resolution Powder Diffractometer (HRPD)
at the ISIS spallation neutron source, Rutherford Appleton Laboratory,
UK.  Data collected on the backscattering (168.33$^\circ$) and
90$^\circ$ detector banks were used.  These data were corrected for
self-shielding and wavelength-dependent absorption for a sample with a
number density of $2.4\times10^{-3}$\,\AA$^{-3}$, scattering cross
section 41.72\,barns, and absorption cross section 36.42\,barns at a
wavelength of 1.798\,\AA.  Powder diffraction data were
Rietveld-refined in FullProf by the least-squares
method\cite{FullProf}.

\section{Crystal Structure}

\begin{figure}[htb]
\includegraphics[width=\columnwidth]{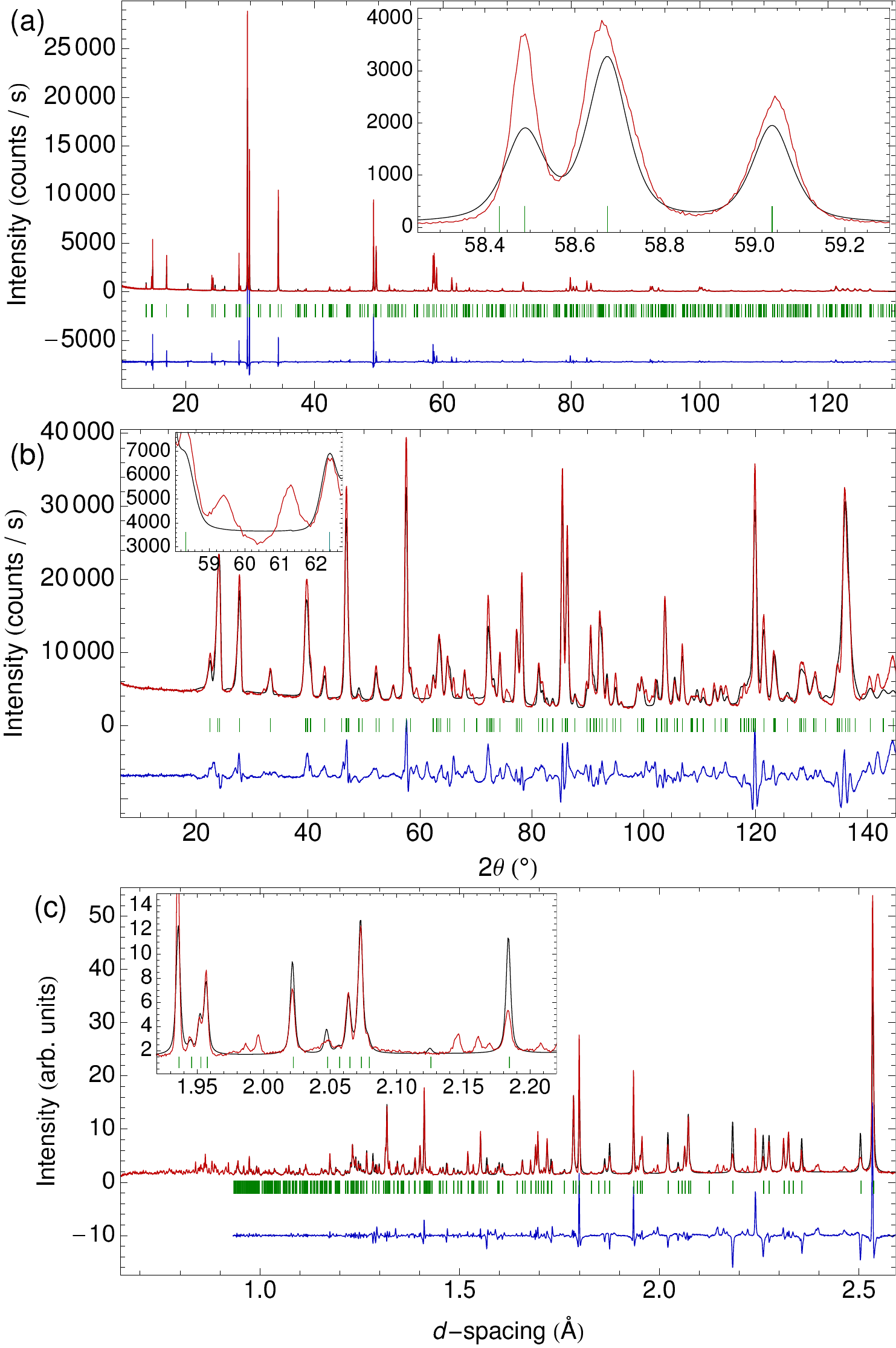}
\caption{\label{fig:diffract}Powder diffraction patterns of \Mn\ at
  300\,K by (a) synchrotron x-ray diffraction, (b) neutron
  diffraction, and (c) time-of-flight neutron diffraction.  Data are
  in red, the result of refinement within the \P3\ structure is
  shown in black, the residual is in blue, and green vertical bars
  mark nuclear Bragg positions.  The insets highlight illustrative
  peaks that cannot be well reproduced within \P3.  The residuals have
  been shifted vertically for clarity.}
\end{figure}

\begin{figure}[htb]
  \includegraphics[width=\columnwidth]{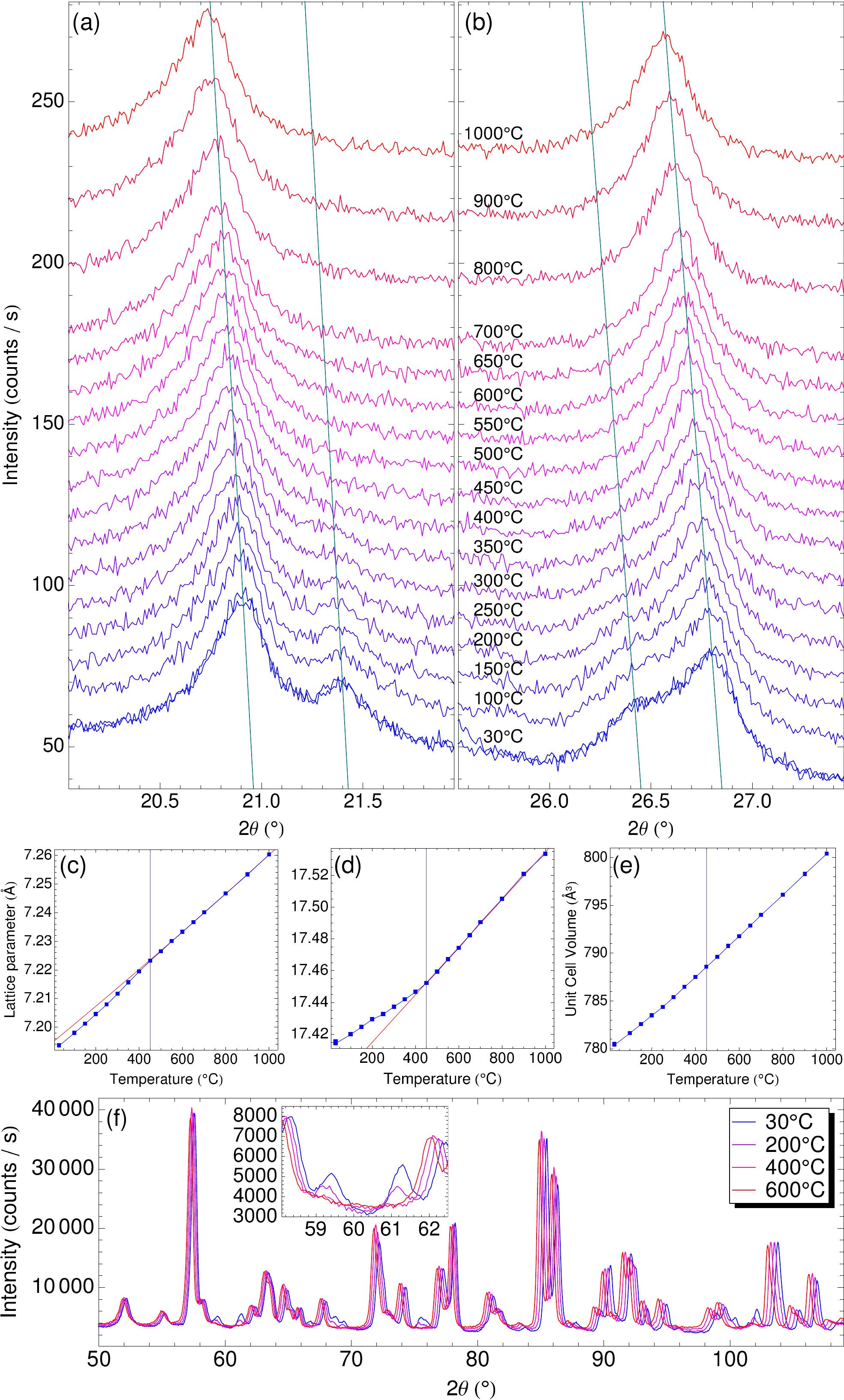}
  \caption{\label{fig:highTP2}Transition to \P3\ at high temperature.
    In the laboratory x-ray data in panels (a) and (b), peaks
    attributable to the $P2$ structure weaken above room temperature,
    becoming indistinct above 350$^\circ$C (data sets have been
    shifted vertically for clarity).  The dual traces at 30$^\circ$C
    were taken first and last, as an internal check.  The (c) $a$-axis
    and (d) $c$-axis lattice parameters show evidence for a transition
    around 450$^\circ$C (marked with a vertical line), but there is no
    clear effect on the (e) unit cell volume.  (f) Selected neutron
    diffraction data: several clear $P2$-derived peaks vanish above
    the structural transition --- two are highlighted in the inset.}
\end{figure}

The crystal structure of \Mn\ has been reported in the \P3\ space
group (\#\,152)\cite{Scott1987,Scott1990,Chelazzi2013}, but there is
also a parenthetical remark in the literature that neutron diffraction
could only be explained satisfactorily using a doubled unit cell in
the $P2$ space group (\#\,3)\cite{Reimers1991}.  Our 300\,K
synchrotron and neutron powder diffraction patterns, shown in
Fig.~\ref{fig:diffract}, indeed show additional features that cannot
be explained within the published \P3\ structure.  The inset to
Fig.~\ref{fig:diffract}(a) shows a set of peaks that demonstrate this
issue in the synchrotron data --- the middle peak is clearly a
doublet, but it corresponds to only a single reflection in \P3, while
the middle and higher peaks are shifted in opposite directions in a
way that cannot be modelled by any unit cell or geometric adjustments
within \P3.  The 300\,K neutron diffraction pattern in
Fig.~\ref{fig:diffract}(b) and the time-of-flight data in
Fig.~\ref{fig:diffract}(c) contain several extra peaks (also see
Fig.~\ref{fig:highTP2}(f)), several of which are shown in the insets.
In addition, modeling of peak intensities is poor in all patterns,
and this is not resolvable through shifts in atomic positions.

Profile matching in all subgroups of \P3\ and various supercells
confirmed that the $P2$ space group with a unit cell doubled in-plane
was required to explain the observed peak positions.  All peaks could
be explained within this $P2$ supercell, and impurity phases could not
explain the additional peaks.  In an independent check, varying the
stoichiometry of the starting materials and the calcining conditions
had no effect on the putative $P2$ reflections, but a several-percent
nonstoichiometry was sufficient to introduce clear impurity peaks at
other angles.  Attempts to refine the $P2$ structure were not
successful --- the presumed $P2$ unit cell would have 70 unique atomic
sites and almost no symmetry operations to constrain them, so a
successful structure refinement will almost certainly require single
crystal diffraction data.  Based on a refinement of the neutron
time-of-flight and synchrotron data with all sites locked to their
ideal \P3\ positions, the $P2$ cell would have lattice parameters
$a=12.46137(45)$\,\AA, $b=7.19304(23)$\AA, $c=17.40822(30)$\AA, and
$\beta=89.9108(20)^\circ$ (neutron), or $a=12.45226(15)$\,\AA,
$b=7.18791(8)$\AA, $c=17.40312(12)$\AA, and $\beta=89.92825(83)^\circ$
(synchrotron) at room temperature.

Since the symmetry reduction to $P2$ was clearer in neutron than in
synchrotron data, the atomic displacements from high-symmetry
positions are likely strongest for oxygen, and one may expect the
structure to return to \P3\ at high temperature.  Diffraction
confirmed this --- peaks associated with $P2$ gradually weakened upon
heating, becoming indistinct around 350$^\circ$C.  Example laboratory
x-ray diffraction patterns are shown in Fig.~\ref{fig:highTP2}(a) and
(b); neutron diffraction results are shown in
Fig.~\ref{fig:highTP2}(f).  The gradual temperature evolution suggests
that this is a second-order displacive transition.  Clear evidence for
the transition may also be observed in the lattice parameters obtained
from refinements within the \P3\ structure, shown in
Fig.~\ref{fig:highTP2}(c) and (d).  A reduction in the $a$-axis and an
increase in the $c$-axis lattice parameters relative to a $\P3$
extrapolation are observed below the transition, estimated from these
data as 450$^\circ$C.  The unit cell volume shows no obvious change
across the transition.  The results of a joint neutron and x-ray
refinement in \P3\ at 600$^\circ$C are included in Fig.~\ref{fig:600}
and Tab.~\ref{tab:600C}.  The shortest Mn--Mn bond lengths are
Mn(1)\,--\,Mn(2), which form a helix along the $c$-axis, while the
kagome-plane triangles are not far from equilateral at this
temperature.

\section{Magnetic Transitions}

\begin{figure*}[htb]
\includegraphics[width=0.85\textwidth]{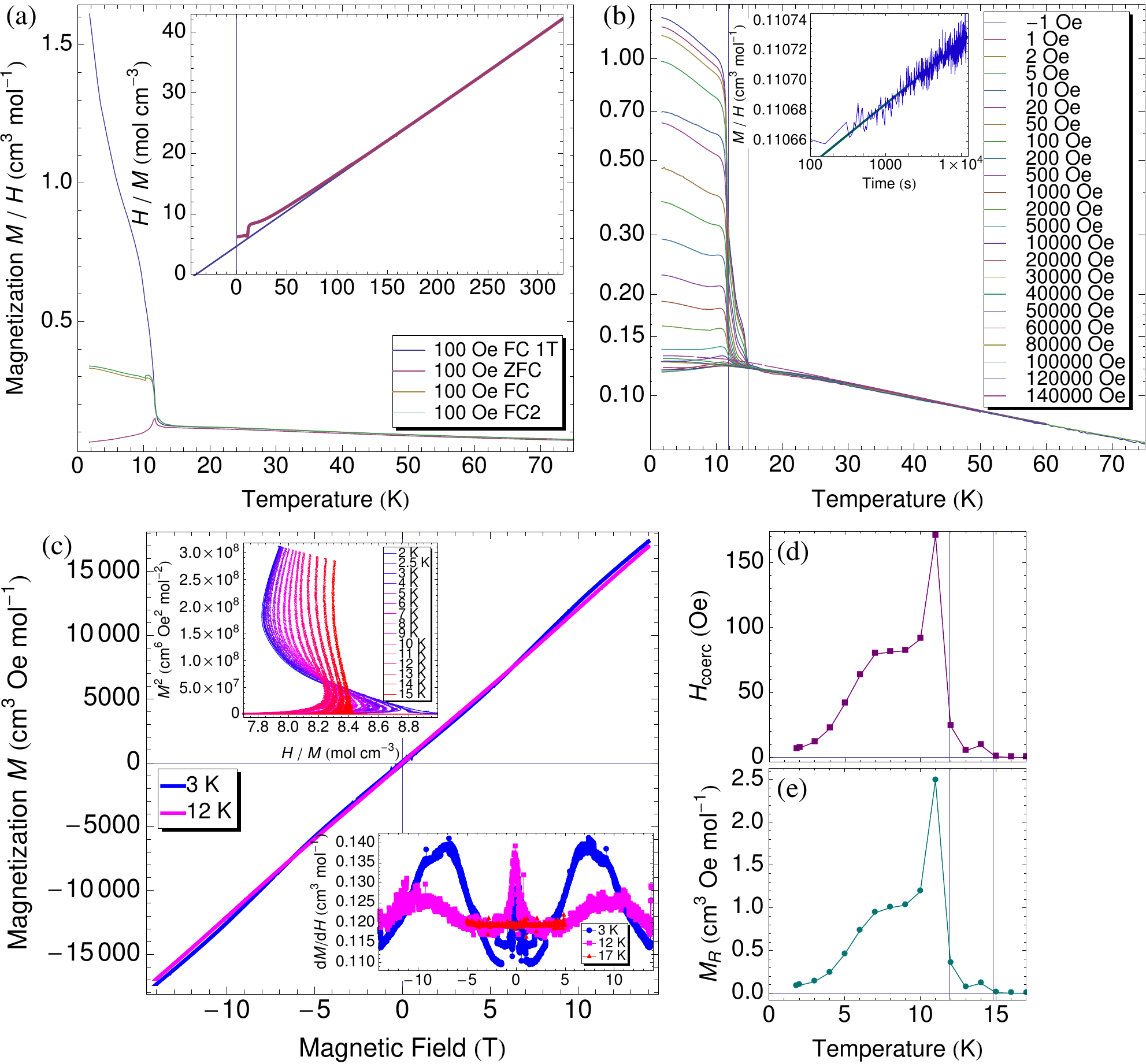}
\caption{\label{fig:MT}Low-temperature magnetization of \Mn.  (a) The
  field-cooled (FC) magnetization $M/H$ in 100\,Oe is compared against
  zero-field cooled (ZFC) data in the same field, and against data
  taken on warming in 100\,Oe after cooling in an applied field of
  1\,T as in Ref.~\onlinecite{Reimers1991}.  The inset shows the
  inverse magnetization (FC, 2000\,Oe), from which a paramagnetic
  moment of 5.88\,$\mu_B$ and a Curie-Weiss temperature of $-40$\,K
  were extracted.  (b) Field-cooled magnetization in a variety of
  fields, plotted on a logarithmic scale.  Vertical lines mark the
  transitions in this sample's specific heat, and the inset shows the
  result of a relaxation measurement at 1.8\,K.  (c) $M$--$H$ loops at
  1.8 and 12.75\,K, with their derivatives in the lower inset.  Below
  the first transition, there is clear curvature and slight
  hysteresis.  The upper inset is an Arrott plot. The hysteresis in
  $M(H)$ can be characterized by (d) the coercive field $H_{coerc}$ or
  (e) the remnant magnetization $M_R$.}
\end{figure*}

Among the few physical properties reported on \Mn\ is
magnetization\cite{Reimers1991}.  An increase in the magnetization was
observed upon cooling through 13\,K, taken to be the bulk ordering
temperature, while the history-dependence below 55\,K was attributed to
the onset of short-range correlations.  This history dependence was
manifested as an approximate doubling of the measured magnetization in
the paramagnetic phase under 100\,Oe if first cooled in a much higher
field\cite{Reimers1991}.  Our preliminary magnetization measurements
indeed detected a strong history-dependent upturn in magnetization on
cooling through a similar temperature range, but this was entirely
attributable to ferrimagnetic Mn$_3$O$_4$ --- the elimination of trace
Mn$_3$O$_4$ impurities eliminated any features or history dependence
above the bulk transition.  The magnetization is plotted in
Fig.~\ref{fig:MT}(b) with a logarithmic vertical axis to enhance any
such weak transition, demonstrating the complete absence of history
dependence.

Fig.~\ref{fig:MT}(a) shows data collected as in
Ref.~\onlinecite{Reimers1991}, as well as zero-field-cooled data.
A history-dependence, while absent above the bulk transition, does
appear at low temperatures, indicating some form of magnetically
frozen state, a weak ferromagnetic component, or possibly magnetic
domains.  This would need to be a rather weak ferromagnetic component,
since much stronger magnetization jumps were observed from trace
ferrimagnetic Mn$_3$O$_4$ that was below the detection limit of
laboratory x-ray diffraction.  The inverse susceptibility, shown in
the inset to Fig.~\ref{fig:MT}(a), departs from linearity below
$\sim$150\,K, indicating short-range correlations appearing well ahead
of any magnetic order.  A fit to the high-temperature inverse
susceptibility returns a Curie-Weiss temperature \TCW\ of $-40$\,K and
a paramagnetic moment of 5.88\,$\mu_B$, consistent with the spin-only
value for $3d^5$ Mn$^{2+}$ of 5.92\,$\mu_B$ and with the previous
report\cite{Reimers1991}.  The onset of short-range correlations well
above the transition and the approximate factor of 3 between the
Curie-Weiss temperature and the magnetic ordering temperature indicate
significant frustration.

The field-cooled magnetization $M(T)$, shown in Fig.~\ref{fig:MT}(a)
and (b), shows a striking increase upon cooling through the bulk
transition, after which it saturates several Kelvin lower.  As will be shown
below, these are two separate transitions.  The field-dependent
magnetization $M(H)$ in Fig.~\ref{fig:MT}(c) is weakly S-shaped with a
change in curvature around 7\,T, and it shows a slight hysteresis,
especially upon entering the $M(T)$ plateau, as can be seen in the
temperature dependence of the coercive field $H_{coerc}$ and remnant
magnetization $M_R$ in Figs.~\ref{fig:MT}(d) and (e), respectively.
The large field dependence in the low-temperature magnetization
$M(T)/H$ at low field, when compared with the minor differences in
$dM/dH$ (lower inset to Fig.~\ref{fig:MT}(c)), reflect the
ferromagnetic component.  Changes of slope in $dM/dH$ at 3\,K around 6
and 9\,T indicate possible metamagnetic transitions, which shift to
{\sl higher} fields in the 12\,K data.  These may have been split or
broadened by powder-averaging of the material's presumably anisotropic
magnetic properties.  That the field scale increases with increasing
temperature is somewhat unusual.  Arrott
plots\cite{Belov1956,Arrott1957,Arrott1967,Bustingorry2016} were also
constructed, as shown in the upper inset in Fig.~\ref{fig:MT}c.  Near
a conventional ferromagnetic transition, these would be linear with a
zero intercept.  The S-shaped curves here serve to accentuate the
changes in curvature in the $M(H)$ data.


\begin{figure}[htb]
  \includegraphics[width=\columnwidth]{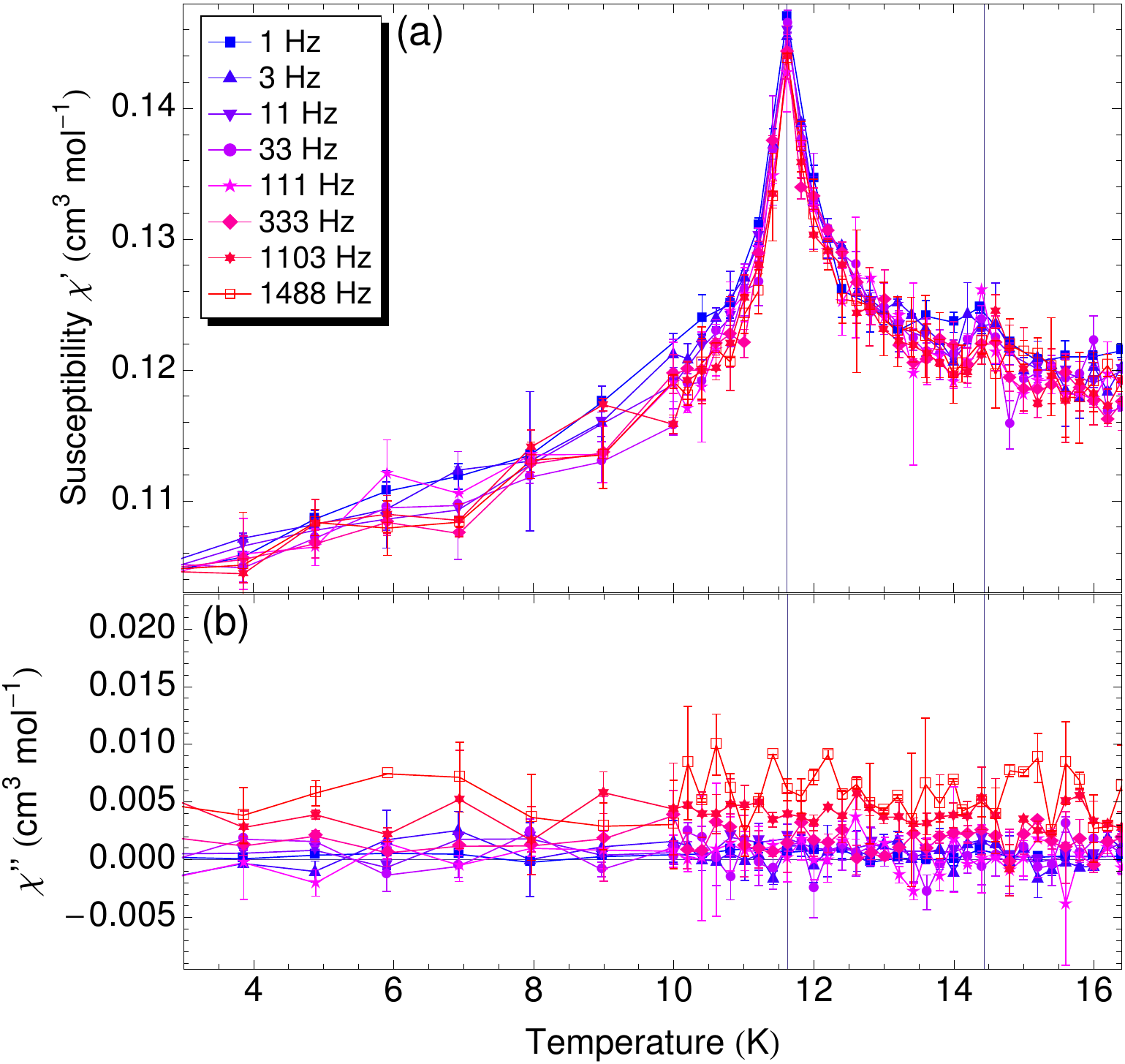}
  \caption{\label{fig:ACchi}AC susceptibility of \Mn.  (a) The real
    component of the AC susceptibility shows features at both
    transitions, while the (b) imaginary component is featureless.  No
    frequency dependence or temperature-dependent dissipation was
    observed.  Error bars represent the statistical uncertainty in the
    fit used to extract each data point.}
\end{figure}

One possible explanation for the shape of the $M(T)$ curves could be a
transition into a spin state with glassy dynamics, although the
temperature dependence of the hysteresis in Figs.~\ref{fig:MT}d and
\ref{fig:MT}e does not take the expected form.  To test for relaxation
behavior, a sample was cooled to 1.8\,K in zero field, the field was
increased to 100\,Oe, and the decay of the magnetization toward its
equilibrium value was then measured.  The extremely weak relaxation
behavior, visible in the inset to Fig.~\ref{fig:MT}(b), is well
described by a single time constant.  Present at the
parts-per-thousand level at best, this may not be intrinsic.  AC
susceptometry was also used to check for glassy behavior at the
transitions (Fig.~\ref{fig:ACchi}): the real component of the
susceptibility, $\chi'$, shows a sharp feature at the lower transition
and a much weaker feature at the upper transition, at temperatures of
11.6 and 14.4\,K respectively.  There is no frequency dependence to
suggest glassy dynamics at the lower transition, while the noise level
precludes any definitive pronouncement as to the nature of the upper
transition.  However, a material would not be expected to enter a
glass state upon cooling and then pass through a subsequent phase
transition into a long-range ordered state --- this would normally be
prevented by the glass states's diverging timescale for relaxation.
The lower panel of this figure, showing $\chi''$, contains no hint of
an onset in dissipation at either transition.

\begin{figure}[htb]
\includegraphics[width=\columnwidth]{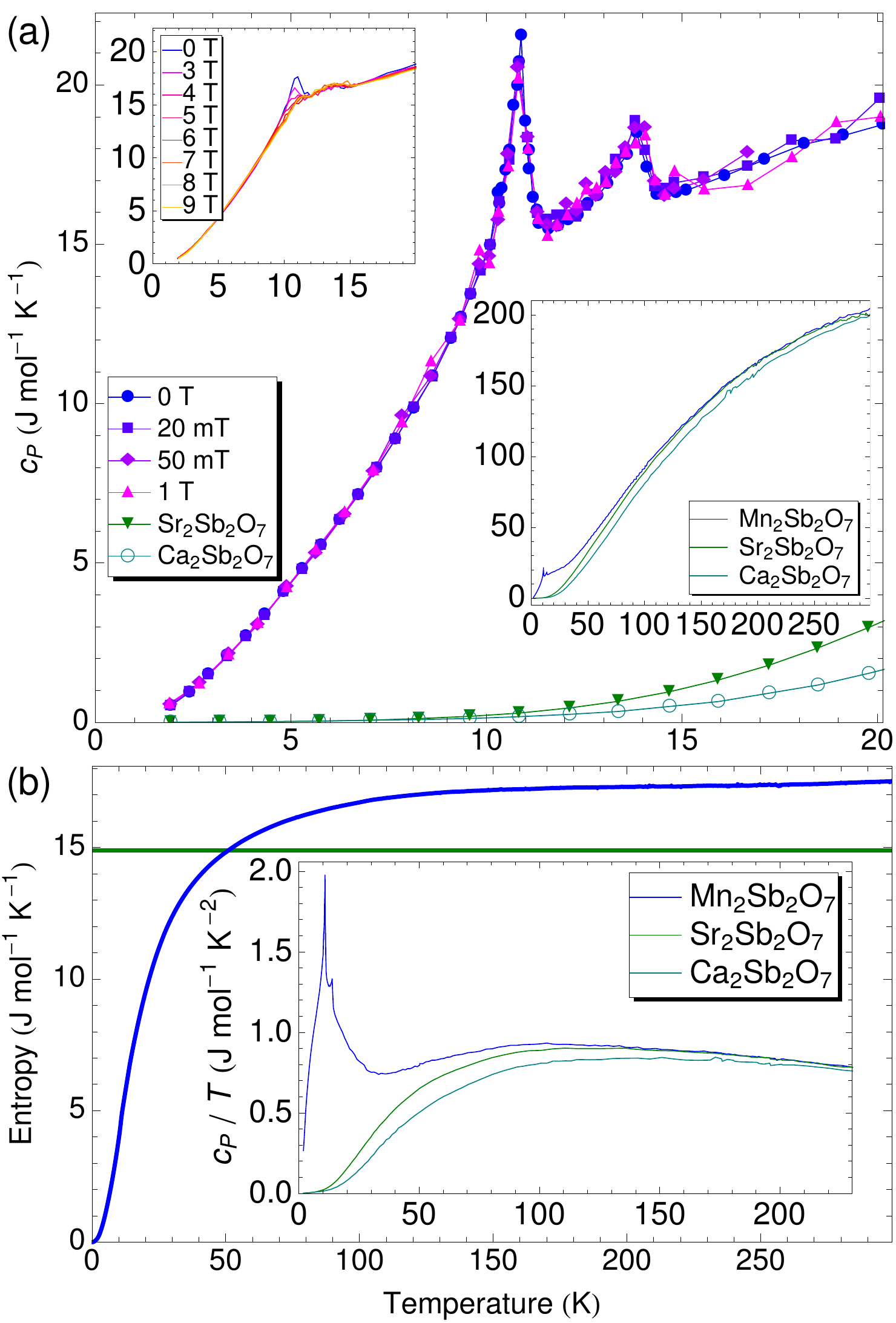}
\caption{\label{fig:cP}(a) Low-temperature specific heat of \Mn,
  showing two phase transitions.  The upper inset shows high-field
  behavior on a second sample, while the lower inset shows the
  high-temperature specific heat at zero field.  Nonmagnetic
  Sr$_2$Sb$_2$O$_7$ and Ca$_2$Sb$_2$O$_7$ are included for comparison.
  (b) The accumulated magnetic entropy in \Mn\ obtained by subtracting
  that of Sr$_2$Sb$_2$O$_7$.  The horizontal line marks the $R\ln 6$
  expected for $s=5/2$; the fact that the entropy exceeds this
  indicates that the phonon subtraction is imperfect.  A plot of
  $c_P/T$ for extracting the magnetic entropy is included in the inset
  --- the phonon contribution in \Mn\ clearly extends lower in
  temperature than in the Sr or Ca analogs.}
\end{figure}

The low-temperature specific heat of \Mn\ is shown in
Fig.~\ref{fig:cP}.  Here, two phase transitions are clearly visible,
at 11.1 and 14.1\,K.  Applied fields up to $\sim$2\,T had no effect,
but the peaks began to broaden noticeably by 5\,T and became
indistinct at higher fields (upper inset).  This is likely due to
powder averaging of an anisotropic field dependence in the phase
transitions.  The specific heat implies that the unusual
step-and-saturation shape in the temperature-dependent magnetization
data actually reflects the onsets of at least two distinct forms of
order.  Marking the specific heat transitions on the $M(T)$ data in
Fig.~\ref{fig:MT}(b) shows that the upper transition corresponds to
the first sudden increase in magnetization, while the lower transition
coincides roughly with the onset of its saturation.


Examining the specific heat to higher temperatures [lower inset to
  Fig.~\ref{fig:cP}(a)], one finds a large build-up of magnetic
entropy below $\sim$40-50\,K.  This onset is clearer when the data are
plotted as $c_P/T$, as in the inset to Fig.~\ref{fig:cP}(b).  Included
for comparison are specific heat data for Ca$_2$Sb$_2$O$_7$ and
Sr$_2$Sb$_2$O$_7$\cite{Knop1980}.  These form in the closely-related
trigonal Weberite structure\cite{Verscharen1978,Cai2009} and are
nonmagnetic insulators.  Sr$_2$Sb$_2$O$_7$ more closely mimics the
high-temperature behavior of \Mn, but its use as a phonon baseline
results in an entropy exceeding the $R\ln6$ expected for $3d^5$
Mn$^{2+}$, as shown in Fig.~\ref{fig:cP}(b).  This is most likely due
to differences in the low-energy phonon spectrum, which would be
unsurprising given the Mn version's relaxation into a lower-symmetry
$P2$ structure.  Mixing of phonons with other types of modes may also
be a possibility in this space group.

\begin{figure}[htb]
  \includegraphics[width=\columnwidth]{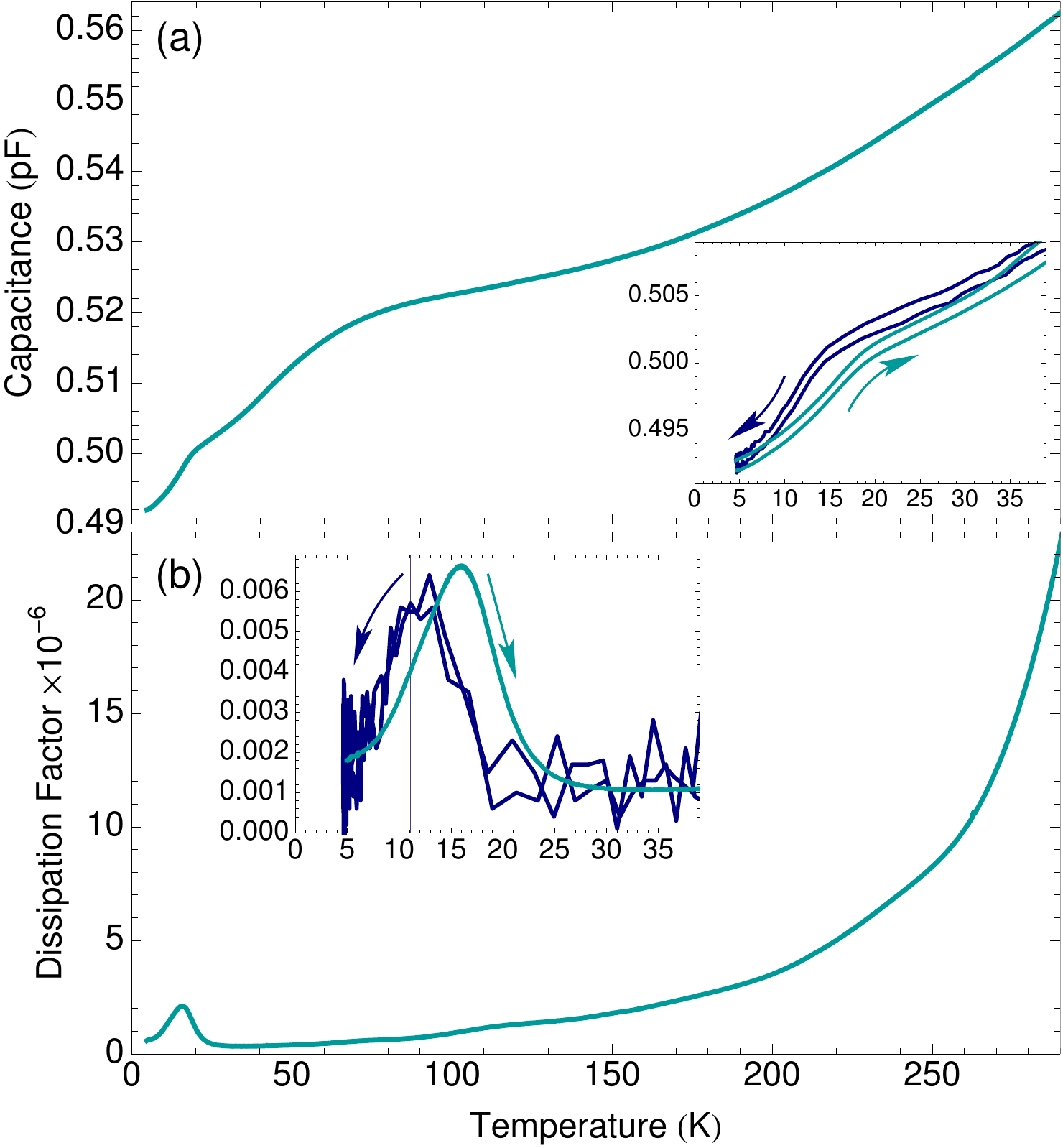}
  \caption{\label{fig:dielectric}Temperature dependence of the (a)
    capacitance and (b) dissipation factor when \Mn\ is employed as
    the dielectric in a capacitor.  The insets show multiple warming
    and more-rapid cooling runs, with the specific heat transitions
    marked for reference.}
\end{figure}

Since the low-temperature $P$2 space group supports ferroelectricity
while the \P3\ space group does not, multiferroicity is a possibility
in this system and measurements of the dielectric properties, shown in
Fig.~\ref{fig:dielectric}, can provide further insight into the phase
transitions.  The capacitance shows a sharp downturn at the upper
magnetic transition, where the magnetic order locking-in reduces the
material's ability to electrically polarize (and store energy as a
dielectric).  A corresponding peak in the dissipation factor is
attributed to order parameter fluctuations near the transition. The
appearance of such fluctuations for a magnetic transition in an
electrostatic quantity implies multi-component order.  The capacitance
undergoes a broader but stronger reduction centered around 50\,K, with
no corresponding peak in the dissipation factor.  This represents the
onset of short-range magnetic order and matches well with the heat
capacity estimate.  Above $\sim$100\,K, where magnetism plays no
significant role, the reduction in the capacitance (and thus
polarizability) upon cooling is suggestive of a tendency toward
antiferroelectric order below the structural transition, in much the
same way that a reduction in magnetic susceptibility is seen below a
pure antiferromagnetic transition.

Data were collected on free warming to minimize noise, and \Mn's
specific heat is very large at low temperatures, so the sample
temperature lags the thermometry leading to an apparent shift of the
data to slightly higher temperatures.  To confirm that the upper
transition is indeed the electrically-active one, data taken upon
cooling with a sweep rate roughly an order of magnitude faster are
included in the insets to Fig.~\ref{fig:dielectric}.  All observed
features remain above the lower transition, unambiguously identifying
the upper transition as the electrically active one.

The clear detection of a magnetic transition and its order parameter
fluctuations in bulk electrostatic properties implies strong
magnetoelectric coupling and mixed-character (multiferroic) order
parameters.  It also confirms that the space group supports
ferroelectricity, lending additional support to the assignment of
$P2$.  

\section{Magnetic structure}

\begin{figure*}[htb]
\includegraphics[width=\textwidth]{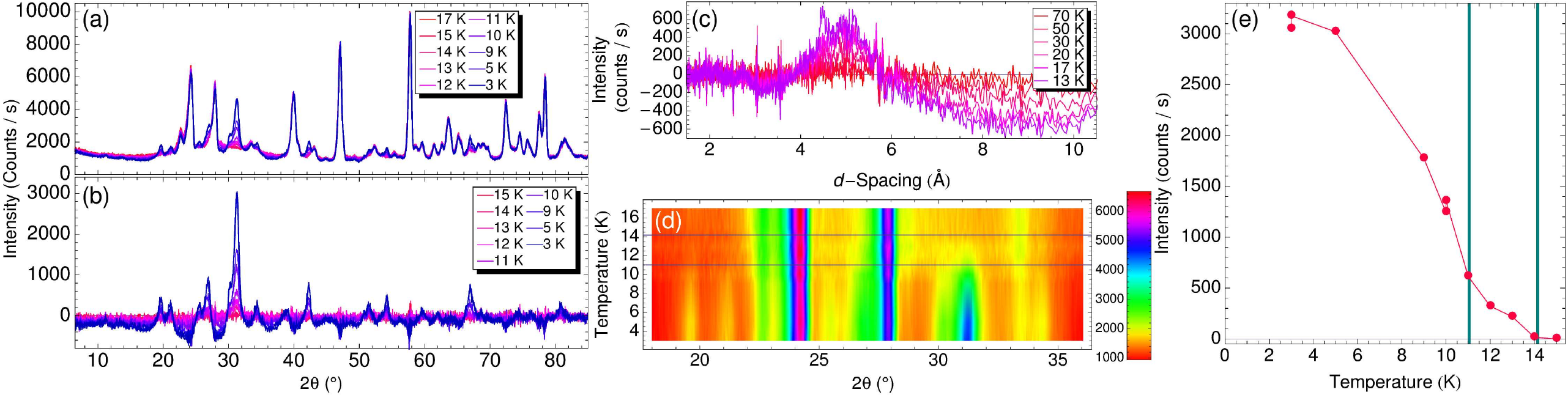}
\caption{\label{fig:neutron-mag}Magnetic intensity at low temperature.
  (a) Diffraction patterns at various temperatures. (b) Difference
  with respect to 17\,K, to highlight the magnetic peaks.  (c) A
  significant diffuse peak is visible above the transition around a
  $d$-spacing of 5\,\AA\ when data taken at 100\,K are used for
  subtraction.  Panels (d) and (e) demonstrate that while much of the
  intensity in these peaks vanishes at the lower magnetic transition,
  some persists to the upper transition.}
\end{figure*}

\begin{table}[htb]
  \caption{\label{tab:mag}Magnetic reflections in \Mn.  Observed and
    calculated $d$-spacings in \AA\ based on data from both ANSTO and
    HANARO, approximate relative intensities (Rel.\ Int.), and index
    in \P3\ and $P2$ of the observed magnetic peaks are listed.}
  \begin{tabular}{cccccr}\hline
    \P3 & $P2$ & $d$ & $d_{HANARO}$ & $d_{calc}$ & Rel.\ Int. \\ \hline\hline
    ($\frac12$00) & (100) & 12.44 & --- & 12.43 & 3.9 \\ \hline
    ($\frac12$01) & (101) & 10.15 & --- & 10.12 & 2.6 \\ \hline
    ($\frac12$02)/($\frac12$10) & (102)/(010) & 7.15 & 7.06 & 7.13 & 27 \\ \hline
    ($\frac12$11) & (011) & 6.65 & 6.55 & 6.64 & 24 \\ \hline
    ($\frac12$12) & (012) & 5.53 & 5.47 & 5.53 & 23 \\ \hline
    ($\frac12$03) & (103) & 5.26 & 5.20 & 5.25 & 39 \\ \hline
    ($\frac32$10) & (210) & 4.68 & 4.65 & 4.70 & 33 \\ \hline
    ($\frac32$11) & (211) & 4.53 & 4.48 & 4.53 & 100 \\ \hline
    ($\frac32$00)/($\frac32$12) & (212) & 4.13 & 4.08 & 4.13 & 24 \\ \hline
    ($\frac32$01) & (301) & 4.03 & --- & 4.03 & 9.8 \\ \hline
    ($\frac12$14)/($\frac32$02) & (014)/(302) & 3.71 & 3.66 & 3.72 & 31 \\ \hline
    ($\frac32$21)/($\frac32$03) & (121)/(303) & 3.38 & 3.36 & 3.38 & 26 \\ \hline
    ($\frac52$10) & (410) & 2.85 & 2.84 & 2.85 & 4.9 \\ \hline
    ($\frac12$06)/($\frac52$11) & (106)/(411) & 2.81 & 2.80 & 2.81 & 11 \\ \hline
    ($\frac12$16)/($\frac52$21) & (016)/(321) & 2.68 & 2.66 & 2.68 & 17 \\ \hline
    several & several & 2.21 & 2.20 &  & 23 \\ \hline
    several & several & 2.20 & 2.19 &  & 6.6 \\ \hline
    several & several & 1.88 & 1.88 &  & 14 \\ \hline
  \end{tabular}
\end{table}

Neutron diffraction was performed through both magnetic transitions
--- data in the relevant temperature range appear in
Fig.~\ref{fig:neutron-mag}(a).  Fig.~\ref{fig:neutron-mag}(b) shows
the same data after subtraction of a pattern taken at 17\,K, just
above the upper magnetic transition.  All magnetic peaks can be
qualitatively explained by a propagation vector of ($\frac12$\,0\,0)
in the \P3\ unit cell, or (0\,0\,0) in the larger $P2$ cell, as
summarized in Tab.~\ref{tab:mag}.  Since the underlying $P2$ crystal
structure could not be refined, and since the number of observed
magnetic peaks is similar to the expected number of Mn sites,
refinement of the magnetic structure was not completed.  The $P2$
space group supports two irreducible representations for the magnetic
order, both of which are chiral, and it was not possible to
distinguish between them.  Calculated $d$-spacings in
Tab.~\ref{tab:mag} are based on the $P2$ assignments.  Based on the
limited data, it appears that peaks with $h+k=$\,odd in $P2$ are
favored.

As has previously been reported\cite{Reimers1991}, there is a
significant diffuse peak centered around a $d$-spacing of 5\,\AA, and
2\,\AA\ wide --- see Fig.~\ref{fig:neutron-mag}(c).  Comparison to 300\,K
data indicated that this feature is essentially absent above 70\,K, so
data collected at 100\,K were used as a baseline for subtraction.  The
appearance of the diffuse peak around 50-70\,K provides further
evidence for the onset of local spin correlations around that
temperature.  Its intensity reaches a maximum between the two magnetic
transitions, and then the magnetic intensity is transferred into the
magnetic Bragg peaks at low temperatures.

The temperature dependence of the magnetic peaks is shown in
Figs.~\ref{fig:neutron-mag}(d) and \ref{fig:neutron-mag}(e).  The
intensity falls off as would be expected for a second-order transition
at 11\,K, but a small fraction of the intensity persists to the higher
magnetic transition.  The magnetic peaks in both phases are located at
the same angles and can be described by the same shape parameters and
intensity ratios, although the low intensity between the two
transitions makes firm conclusions difficult.  The strong similarity
indicates that the two magnetic phases are closely related --- perhaps
distinguished by a canting angle, stacking along the hexagonal $c$
axis, or coupling to antiferroelectric order.

\section{Discussion and Conclusion}

As is clear from the magnetization, specific heat and neutron
diffraction data, \Mn\ undergoes two magnetic phase transitions at low
temperature.  The upper transition is multiferroic in nature, while
the lower is purely magnetic.  Below the lower transition, the
magnetization saturates and the hysteresis reaches its maximum before
subsiding, suggesting that the ground state is some form of global
antiferromagnetic order.  Between the two transitions, the
magnetization increases sharply upon cooling, implying a small net
moment.  The temperature-dependence of the apparent high-field
transitions in $M(H)$ suggests frustration is relieved by the field.

The magnetic sublattice is intermediate between two-dimensional kagome
and three-dimensional pyrochlore, so this material offers a platform
for interpolating between two- and three-dimensional frustrated
magnetism.  The clearest comparison to a pyrochlore system is to the
pyrochlore polymorph of
\Mn\ itself\cite{Brisse1972,Zhou2008,Zhou2010,Peets-pyr}.  This latter
structural variant has only been successfully prepared {\it via} a
specific low-temperature route, substituting Mn$^{2+}$ into the
pre-existing pyrochlore Sb$^{5+}$ framework of Sb$_2$O$_5\cdot
n$H$_2$O ($n\sim 1.5$\,--\,2).  By having 6 Mn--Mn links within a
tetrahedron rather than five within an armchair unit, one might expect
the pyrochlore's Curie-Weiss temperature to be 20\%\ stronger, or
$-48$\,K.  Experimentally, we found it to be $-49$\,K\cite{Peets-pyr}.
One would also expect a significantly higher frustration factor
$f\equiv \TCW/T_\text{N}$ in the pyrochlore, and again this is exactly
what we observe: $f_\text{ak} = 2.8$ for the armchair-kagome structure
and $f_\text{pyr} = 9.0$ for the pyrochlore.  The stronger frustration
in the pyrochlore polymorph leads to a spin-glass ground state, rather
than the three-dimensional magnetic order observed here.

\begin{figure}[htb]
  \includegraphics[width=\columnwidth]{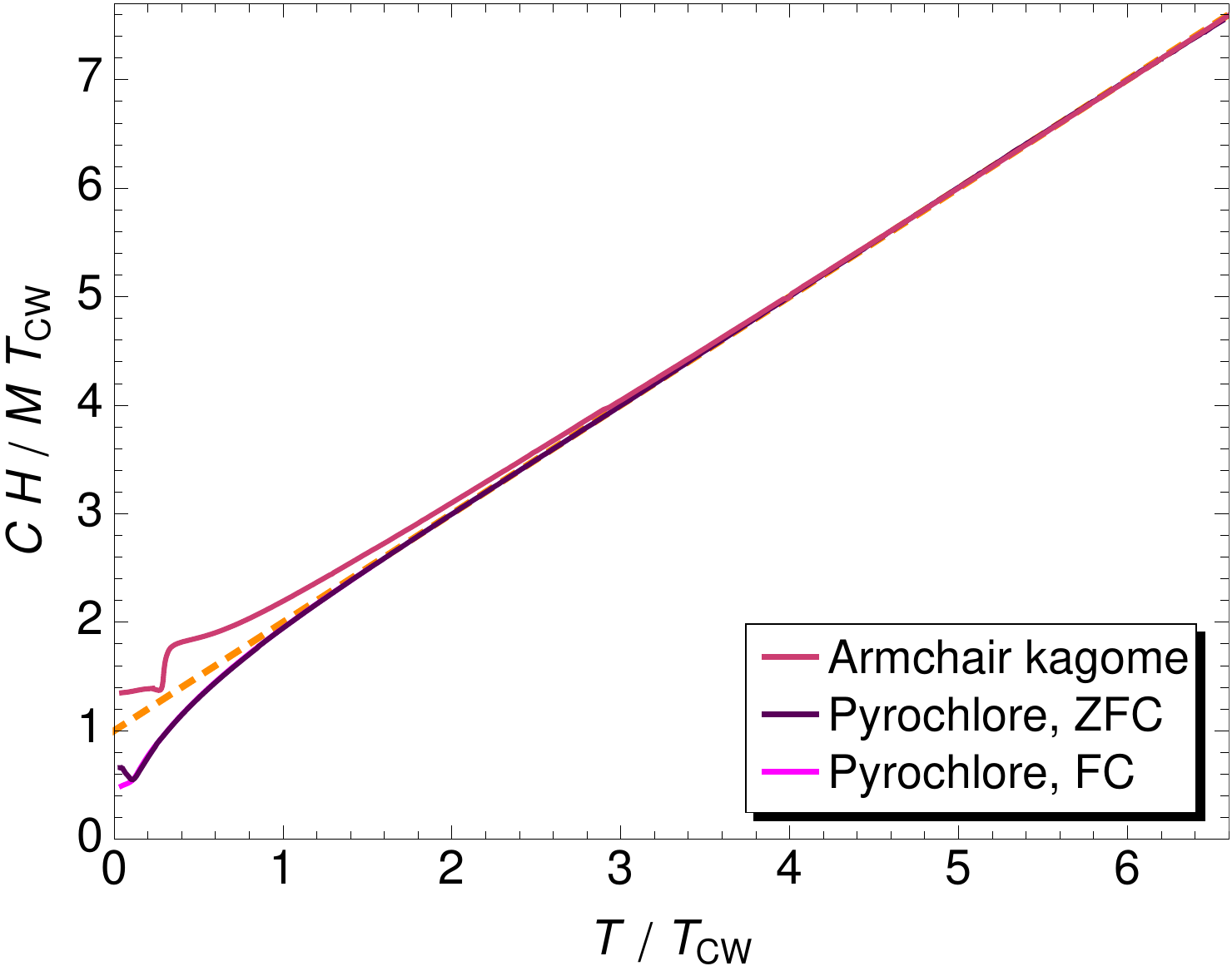}
  \caption{\label{compare}Comparison of the armchair-kagome and
    pyrochlore variants of \Mn.  Magnetization data on the former
    polymorph were taken at 2000\,Oe (field-cooled), and are the same
    data as in the Fig.~\ref{fig:MT}a inset.  Data on the pyrochlore
    were taken at 100\,Oe under field-cooled and zero-field-cooled
    conditions\cite{Peets-pyr}.}
\end{figure}

For a more direct comparison, temperature-dependent magnetization data
are plotted in unitless form for both polymorphs in
Fig.~\ref{compare}, obtained by rearranging the Curie-Weiss law $M/H =
C/(T-\TCW)$ as $CH/M\TCW = T/\TCW -1$\cite{Dutton2011}.  Data above
the dashed line indicate additional antiferromagnetic interactions,
while points can be pushed below the line by ferromagnetic
interactions.  Deviations from ideal Curie-Weiss behavior start at a
higher fraction of \TCW\ in the armchair-kagome structure and indicate
additional antiferromagnetic fluctuations, whereas the deviations from
Curie-Weiss behavior in the pyrochlore are ferromagnetic.  The
transitions in both polymorphs lie well below the unfrustrated
$T_\text{N}=\TCW$.

In summary, we have shown that the structure of \Mn\ must be $P2$
under ambient conditions, but that it returns to the higher-symmetry
\P3\ structure above $\sim$450$^\circ$C in an apparent second-order
displacive transition.  Solving the $P2$ modification will most likely
require single crystals.  The armchair-kagome network seen in the
\P3\ structure has not, to our knowledge, been modeled theoretically,
and investigation of its possible magnetic and multiferroic ground
states as a function of interaction strengths would be an interesting
and fruitful topic for future research.  The Mn$_4$ armchair units
constitute an intermediate case between the two-dimensional triangular
network of the kagome lattice and the fully three-dimensional
tetrahedra of the pyrochlore, and likely support their own unique
suite of ground states as a function of the exchange parameters.
Interestingly, the smaller Sb$^{5+}$ ions form identical structural
armchair motifs, but with an additional twist between layers that is
not present in the Mn sublattice.  It would be worthwhile to explore
the possibility of putting magnetic ions on this site instead.

Magnetic Bragg peaks are consistent with a propagation vector of
$(\frac{1}{2} 0 0)$ in \P3\ or $(000)$ in $P2$, but a magnetic
structure could not be refined.  Because the latter crystal structure
only supports chiral irreducible representations for the magnetic
order, we infer that the magnetic order in the material is chiral.
Chiral magnetism can lead to a variety of exotic physics, and will be
of significant interest for its excitations and field-dependence.  The
possibility of skyrmion-like excitations\cite{Skyrmions} in a system
with chiral multiferroic order would be particularly enticing, as it
would enable control and manipulation of the magnetism and excitations
through standard electronic means.  In the closely-related
MnSb$_2$O$_6$\cite{Reimers1989}, the novel cycloidal magnetic order
has been predicted to lead to a unique ferroelectric switching
mechanism, while the material should behave in an analogous way to
ferroaxial multiferroics\cite{Johnson2013}, and it would be
interesting to determine whether similar physics could be available
here.  \Mn\ has some key differences, however, and may host its own
suite of entirely unique physics.

\section*{Acknowledgements}
This work was supported by the Institute for Basic Science (IBS) in
Korea (IBS-R009-G1), and work at HANARO was supported by the Nuclear
R\&D Program through NRF Grant No.\ 2012M2A2A6002461.  The authors are
indebted to M.\ Gingras, D.I.\ Khomskii, M.J.\ Lawler, M.D.\ Le, and
Y.\ Noda for stimulating discussions, the NCIRF for assistance with
several measurements, and K.S.\ Knight at ISIS for assistance with the
time-of-flight diffraction measurement.  We acknowledge the support of
the Bragg Institute, Australian Nuclear Science and Technology
Organisation, in providing neutron research facilities used in this
work.

\bibliography{Mn2Sb2O7,Arrott}

\appendix
\clearpage
\section{High-Temperature {\slshape P}3$_1$21 Structure Refinement}

\begin{figure}[htb]
  \includegraphics[width=\columnwidth]{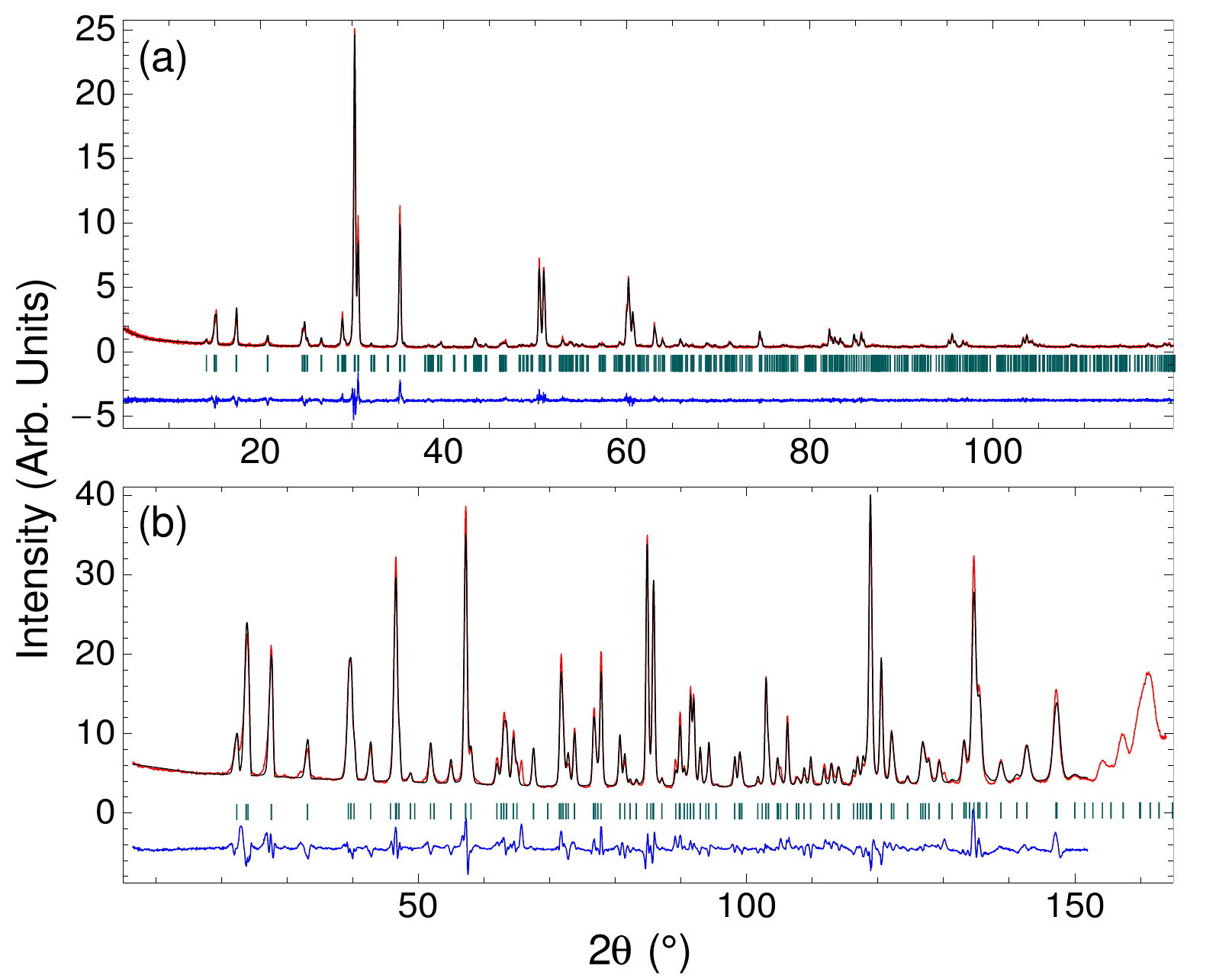}
  \caption{\label{fig:600}Joint refinement of (a) x-ray
    and (b) neutron powder diffraction data at 600$^\circ$C in the
    \P3\ space group (\#\,152).  Data are in red, the fit is in black,
    the residual is in blue, and vertical bars mark the calculated
    Bragg positions.  The residuals have been shifted for clarity.}
\end{figure}



\begin{table}[htb]
  \caption{\label{tab:600C}Refined atomic positions in the \P3\ space
    group (\#\,152) at 600$^\circ$C, well above the structural
    transition, from a joint refinement of x-ray and neutron powder
    diffraction patterns.  $a=7.22277(7)$\,\AA\ and
    $c=17.4479(2)$\,\AA\ in the neutron pattern and
    $a=7.2248(2)$\,\AA\ and $c=17.4545(7)$\,\AA\ in the x-ray pattern.
    The overall $R$-factors based on all points and not corrected for
    background are $R_p=5.92$\%\ and $R_{wp}=7.88$\%\ for the neutron
    pattern and $R_p=8.84$\%\ and $R_{wp}=11.6$\%\ for the x-ray
    pattern, based on 202 neutron reflections and 1199/2 reflections
    for the x-ray pattern.}
  \begin{tabular}{lcr@{.}llr@{.}ll}\hline
    Atom & Mult & \multicolumn{2}{l}{~$x$} & $y$ & \multicolumn{2}{l}{~$z$} & $B_\text{iso}$ \\ \hline\hline
    Mn(1) & 3a & 0&853(4) & 0 & 0&33333 & 3.09(18) \\
    Mn(2) & 3b & 0&868(3) & 0 & 0&83333 & 3.09(18) \\
    Mn(3) & 6c & 0&659(3) & 0.152(4) & -0&0039(8) & 3.09(18) \\
    Sb(1) & 3a & 0&3227(16) & 0 & 0&33333 & 0.78(10) \\
    Sb(2) & 3b & 0&3387(19) & 0 & 0&83333 & 0.78(10) \\
    Sb(3) & 6c & 0&4953(22) & 0.3361(20) & 0&1646(5) & 0.78(10) \\
    O(1) & 6c & 0&1954(17) & 0.2235(14) & 0&1408(5) & 1.78(7) \\
    O(2) & 6c & 0&5617(10) & 0.6161(15) & 0&1987(5) & 1.78(7) \\
    O(3) & 6c & 0&1905(16) & 0.6366(16) & 0&1437(5) & 1.78(7) \\
    O(4) & 6c & -0&0446(12) & 0.3105(18) & 0&0540(3) & 1.78(7) \\
    O(5) & 6c & -0&0396(10) & 0.8045(20) & 0&0555(4) & 1.78(7) \\
    O(6) & 6c & 0&5395(21) & 0.4016(12) & 0&0554(4) & 1.78(7) \\
    O(7) & 6c & 0&5499(15) & 0.8179(20) & 0&0597(4) & 1.78(7) \\ \hline
  \end{tabular}
\end{table}

Figure \ref{fig:600} and Tab.\ \ref{tab:600C} report the result of a
joint refinement of x-ray and neutron powder diffraction data in the
high-temperature \P3\ structure (space group \#\,152) at 600$^\circ$C.
$B_\text{iso}$ parameters were constrained to be the same on all sites for
each element.  The lattice parameters at this temperature were found to
be $a=7.22277(7)$\,\AA\ and $c=17.4479(2)$\,\AA\ in the neutron
pattern and $a=7.2248(2)$\,\AA\ and $c=17.4545(7)$\,\AA\ in the x-ray
pattern.  The overall $R$-factors based on all points and not
corrected for background were $R_p=5.91$\%\ and $R_{wp}=7.87$\%\ for
the neutron pattern and $R_p=8.58$\%\ and $R_{wp}=11.3$\%\ for the
x-ray pattern, based on 202 neutron reflections and 1199/2 reflections
in the x-ray pattern.  The resulting Mn--Mn bond lengths and bond
angles are summarized in Tab.\ \ref{tab:P3bond}.  At least at this
temperature, the kagome planes are much closer to their ideal shape
than in the previously-reported structure refinement\cite{Scott1990}.
It is not clear whether this is due to the elevated temperature, the
sample quality, or the previous use of the high-temperature space
group at room temperature.
    
\begin{table}[htbp]
  \caption{\label{tab:P3bond}Mn--Mn distances and Mn--Mn--Mn angles in
    \P3\ \Mn\ (space group \#\,152) at 600$^\circ$C, for the
    refinement reported in Tab.~\ref{tab:600C} and Fig.~\ref{fig:600}.
    Mn(2) is the site linking adjacent kagome planes, Mn(1) has bonds
    to two out-of-plane Mn(2) atoms, and the two Mn(3) atoms each have
    a bond to only one out-of-plane Mn(2) atom.}
  
  \begin{tabular}{cccr@{.}lr@{.}l}\hline
    Atom 1 & Atom 2 & Atom 3 &
    \multicolumn{2}{c}{Bond length (\AA)} &
    \multicolumn{2}{c}{Angle ($^\circ$)} \\ \hline
    
    Mn(1) & Mn(2) & & ~~~~~3&393(16) & \multicolumn{2}{c}{ } \\
    Mn(1) & Mn(3) & & 3&54(3) & \multicolumn{2}{c}{ } \\
    Mn(1) & Mn(3) & & 3&68(2) & \multicolumn{2}{c}{ } \\
    Mn(2) & Mn(3) & & 3&64(2) & \multicolumn{2}{c}{ } \\
    Mn(3) & Mn(3) & & 3&62(3) & \multicolumn{2}{c}{ } \\ \hline
    
    Mn(2) & Mn(1) & Mn(2) & \multicolumn{2}{c}{ } & 126&1(4)\\
    Mn(2) & Mn(1) & Mn(3) & \multicolumn{2}{c}{ } & 63&2(7)\\
    Mn(2) & Mn(1) & Mn(3) & \multicolumn{2}{c}{ } & 115&1(10)\\
    Mn(2) & Mn(1) & Mn(3) & \multicolumn{2}{c}{ } & 92&1(7)\\
    Mn(2) & Mn(1) & Mn(3) & \multicolumn{2}{c}{ } & 90&3(7)\\
    Mn(3) & Mn(1) & Mn(3) & \multicolumn{2}{c}{ } & 177&6(16)\\
    Mn(3) & Mn(1) & Mn(3) & \multicolumn{2}{c}{ } & 60&0(9)\\
    Mn(3) & Mn(1) & Mn(3) & \multicolumn{2}{c}{ } & 121&0(12)\\
    Mn(3) & Mn(1) & Mn(3) & \multicolumn{2}{c}{ } & 119&0(11)\\
    
    Mn(1) & Mn(2) & Mn(1) & \multicolumn{2}{c}{ } & 128&1(4)\\
    Mn(1) & Mn(2) & Mn(3) & \multicolumn{2}{c}{ } & 60&5(8)\\
    Mn(1) & Mn(2) & Mn(3) & \multicolumn{2}{c}{ } & 171&4(8)\\
    Mn(3) & Mn(2) & Mn(3) & \multicolumn{2}{c}{ } & 111&0(8)\\
    
    Mn(1) & Mn(3) & Mn(1) & \multicolumn{2}{c}{ } & 177&6(16)\\
    Mn(1) & Mn(3) & Mn(2) & \multicolumn{2}{c}{ } & 125&3(11)\\
    Mn(1) & Mn(3) & Mn(2) & \multicolumn{2}{c}{ } & 56&4(8)\\
    Mn(1) & Mn(3) & Mn(3) & \multicolumn{2}{c}{ } & 58&1(9)\\
    Mn(1) & Mn(3) & Mn(3) & \multicolumn{2}{c}{ } & 119&0(11)\\
    Mn(1) & Mn(3) & Mn(3) & \multicolumn{2}{c}{ } & 120&8(12)\\
    Mn(1) & Mn(3) & Mn(3) & \multicolumn{2}{c}{ } & 61&9(9)\\
    Mn(2) & Mn(3) & Mn(3) & \multicolumn{2}{c}{ } & 95&8(8)\\
    Mn(2) & Mn(3) & Mn(3) & \multicolumn{2}{c}{ } & 89&3(8)\\
    Mn(3) & Mn(3) & Mn(3) & \multicolumn{2}{c}{ } & 174&9(10)\\ \hline
  \end{tabular}
\end{table}


\end{document}